# Ideal strength of two-dimensional stanene may reach or exceed Griffith strength estimate

Zhe Shi[a], Chandra Veer Singh[a,b] *

[a]Department of Materials Science and Engineering, University of Toronto, Toronto, ON, M5S 3E4, Canada

[b]Department of Mechanical and Industrial Engineering, University of Toronto, Toronto, ON, M5S 3G8, Canada

*Corresponding author

*Abstract*

The ideal strength is the maximum stress a material can withstand, and it is an important intrinsic property for structural applications. Griffith strength limit ~E/9 is the best known upper bound of this property for a material loaded in tension. Here we report that stanene, a recently fabricated two-dimensional material, could approach this limit from a theoretical perspective. Utilizing first-principles density functional theory, we investigated the nonlinear elastic behavior of stanene and found that its strength could reach ~E/7.4 under uniaxial tension in both armchair and zigzag directions without incurring phonon instability or mechanical failure. The unique mechanical properties of stanene are appreciated by comparisons with other Group-IV 2D materials.

## 1. INTRODUCTION

The structural application of any nanostructured materials hinges on a better understanding of their strength and mechanical behavior[1,2]. The ideal (theoretical) strength, $\sigma_{ideal}$, is the maximum stress achievable in a perfect crystal at zero Kelvin[3–5]. Knowledge of this value is important to our understanding of many problems in the solid state, as it essentially corresponds to the failure of a crystal loaded in a defined mode and is crucial to characterize the materials failure which is usually controlled by nucleation and motion of defects. In 1921, Griffith[6] first experimentally extrapolated a theoretical strength of ~E/9 applicable to solid, where E is the elastic modulus of the material under uniaxial tension. Later, Polanyi[7], Orowan[8], and other scientists[9,10], by relating the ideal strength to macroscopic physical properties, set up a similar upper limit of $\sigma_{ideal} \approx E/10$.

Most three-dimensional engineering materials, nonetheless, have an observable (realistic) strength that is many orders of magnitude smaller than their theoretical strength estimated by the above relations, as a significant amount of flaws and defect structures undermines the usable strength. This problem can be greatly subdued if the material is to be made extremely thin or flat at the nanoscale, with the underlying concept that the probability of a critical flaw will reduce as size of the material decreases. The 'idea of thinness' hastened the way to use nanowires[2,11–13], nanopillars[14,15], nanotubes[16], and even nanoropes[17] to achieve ultra-high strength, whereas the 'idea of flatness' put 2D materials onto the stage. Extensive studies have been given to a broad category of 2D materials and satisfying mechanical properties have been obtained. For instance, experimental measurement by atomic force microscope has demonstrated a strength range of E/10 to E/15 for monolayer $MoS_2$ [18]. Previous density functional theory (DFT) simulations echo the experimental result for $MoS_2$ [19] and predict a strength level of ~E/13 [20] for borophene and ~E/11 [21,22] for hexagonal boron nitride and g-GeC. As the strongest material known so far, the study of graphene never falls short. It has been shown that graphene, devoid of any defect, could reach the strength of E/11 [23,24] to E/9 [1] during uniaxial tension. Also, the same Group-IV 2D material silicene has a strength up to ~E/10[25]. These 2D materials all seem to get close to the Griffith limit[6]. The question thus arises is whether a material could possibly surpass these seemingly upper limits on theoretical tensile strength.

The hunt for such candidate has hit the lower part of the Group-IV elements. Ab initio study of the ideal strength of bulk diamond, silicon, and germanium suggests that the decrease of stiffness and strength downward the Group-IV is simultaneous but not proportional[26]. It has been shown that the Young's modulus drops faster than the strength of the material, which results in a reduction of the $E/\sigma_{ideal}$ ratio from diamond to germanium[26]. Starting from this insight, it is very natural to expect the next coming Group-IV 2D material stanene to get further closer to the E/9~E/10 strength limit imposed by Griffith and others on macroscopic solids.

In the 2D materials family, stanene has drawn a particular interest due to its exceptional performances and properties such as topological superconductivity[27], quantum thermal transport[28], quantum anomalous Hall effect[29], and catalytic activity[30]. Recently, successful fabrications of stanene both on substrates[31,32] and as a free-standing structure[33] have been achieved. However, the evidence of stanene sustaining significant mechanical deformation remains elusive. In practical applications, a device made of 2D materials must be able to maintain its mechanical integrity during every stage of its production and function life. Being the strongest material ever produced, graphene seldom 'worries' about its candidacy for applications that require high strength or stiffness. But for an allotrope of tin, softer than carbon in its bulk size, to replace graphene in certain applications, its mechanical robustness has to be assured first. After all, all of those novel properties of stanene cannot be utilized in practice should it fail too easily.

Therefore, investigating the mechanical properties of stanene is important both from a fundamental perspective in understanding its deformational behavior and from a practical interest for its real-world applications. In this paper, we study the mechanical response of stanene at considerable strains and adopt a rigorous continuum formulation to determine the nonlinear elastic constants up to fifth order under uniaxial and biaxial tension. For each deformation case, we determine the ultimate tensile strength (UTS), which represents the highest point on the stress-strain curve and after which the material is considered mechanically failed. Since 2D materials could fail by phonon instability before mechanical failure, the integrity of stanene is also verified by looking at any imaginary phonon frequencies at various strain levels and loading directions. We also compare the properties of stanene with its Group-IV lighter cousins, graphene and silicene, and reveal the possible breakdown of the Griffith theoretical strength limit in stanene. Lastly, charge density and computed scanning tunneling microscopy (STM) images are analyzed to capture the salient features of the deformation and fracture process of stanene.

## 2. MODELING AND THEORY

The stress-strain response of 2D stanene was investigated using first principles based density functional theory (DFT) calculations as implemented in the Quantum ESPRESSO simulation package[34]. In order to confirm the accuracy of quantum chemistry computations, we performed simulations by using the following potentials/functionals: Rappe-Rabe-Kaxiras-Joannopoulos (RRKJ) potential[35] and projector-augmented-wave (PAW) potential both with an exchange-correlation functional of the PBEsol generalized gradient approximation, Martins-Troullier (MT)[36] potential with a functional of the Perdew-Wang (PW) local density approximation (LDA) and with hybrid functionals of PBE0 and B3LYP, and Goedecker-Hartwigsen-Hutter-Teter (GHHT) potential[37] with a hybrid functional of HSE06. The calculations used a kinetic energy cutoff of 816 eV and an $11 \times 11 \times 3$ Monkhorst-Pack k-grid. A force convergence criterion of 0.001 eV/Å was selected for structural optimization. Spin-orbit coupling was shown not to have a distinguishable influence on pure mechanical property calculations in our case (Supplementary Materials Figure S1) and therefore not turned on for all the cases. The monolayer tin was initially constructed using the experimental lattice constant $a = 4.383$ Å [31] and buckled distance $\delta = 1.2$ Å [31]. A vacuum layer of 30 Å was included to reflect the 2D nature of stanene. Simulations were carried out on six-atom supercell for PBEsol, LDA, PBE0 functionals and on two-atom supercell for B3LYP and HSE06 functionals due to high computational expense required by hybrid functionals. The comparative analysis presented in Table S1 and Table S3 shows that the both choices for number of atoms per supercell yield very similar structural and stress behavior, confirming that both systems represent the identical 2D lattice within simulation errors.

The material system was first subjected to a variable cell relaxation to obtain a fully relaxed undeformed configuration. The ground state lattice constant after relaxation was 4.547 Å, within 2% of the values obtained by previous studies[38,39]. The thickness was measured to be 3.26 Å, consistent with the experimental value for free-standing stanene (3.3 Å [33]). The relaxation gave an average Sn–Sn bond length of 2.74 Å, a buckled distance of 0.79 Å, and a bond angle of 111.91° (Figure 1, Table S1). The strains were imposed by specifying the positions of the atoms within the supercell, followed by relaxing the Sn atoms into their equilibrium positions within the deformed structure that yielded the minimum total energy. The stanene was strained uniaxially along the armchair (X) and the zigzag (Y) direction, respectively, or equibiaxially along both directions.

The DFT calculation yielded the 3D Cauchy (true) stress, which was converted to 2D stress with a unit of N/m by taking the product of the stress (N/m$^2$) and thickness of the free-standing stanene. Wei et al's continuum formulation[23] was followed to describe materials elastic response. In order to obtain nonlinear elastic properties from stress-strain relationships derived from elastic strain energy density function, the 2D Cauchy stress ($\sigma$) was related to the second Piola-Kirchoff (PK2) stress $\Sigma$ (N/m) by the following relation[23,40]:

$$\Sigma = JF^{-1}\sigma(F^{-1})^T,$$

where $J$ is the determinant of the deformation gradient tensor $F$. The elastic properties of the material could be determined from the PK2 stress tensor according to $\Sigma_i = C_{ij}\eta_j + \frac{1}{2!}C_{ijk}\eta_j\eta_k + \frac{1}{3!}C_{ijkl}\eta_j\eta_k\eta_l + \frac{1}{4!}C_{ijklm}\eta_j\eta_k\eta_l\eta_m + \cdots$ [23], where η is the Lagrangian strain and the summation convention for the subscripts runs from 1 to 6 employing the Voigt notation[41].

To examine the vibrational stability of the deformed stanene, we calculated the phonon dispersions for stanene based on density functional perturbation theory (DFPT)[42]. We used a two-atom supercell (Figure 1c, d) and a dense 21×21×1 k-grid with a 5×5×1 q-grid to map out possible instabilities. Structural optimization yielded a monolayer structure identical to the six-atom supercell construction (See Table S3 for more information). The PW-LDA pseudopotential was selected for the DFPT calculations, and the same convergence criteria as the stress-strain calculations were kept.

### 3. RESULTS AND DISCUSSIONS

#### 3.1 Stress-strain Response and High-order Elastic Constants

The nonlinear elastic response of stanene in terms of PK2 Stress vs. Lagrangian strain is shown in Figure 2a. Stanene has an isotropic elastic response at strains up to ~10%, evidenced by a coincidence of the stress-strain curves. Compared to armchair tension, stanene is somewhat stronger in the zigzag direction, with the maximum PK2 stress of 3.071 N/m, or 3.656 N/m in true stress. Elongation to UTS for the two loading directions also varies. In the case of zigzag loading, it was found that stanene could sustain approximately 10% more deformation before reaching the peak stress than armchair loading. It is also interesting to note that, the biaxial stress in Figure 2a becomes much higher than the two uniaxial stress responses, whereas the true stress measure in Figure 2b suggests the opposite. This phenomenon reveals the important role of selecting the reference area when analyzing the nonlinear mechanical response.

Higher-order elastic constants for stanene are extracted based on Wei et al's the thermodynamically rigorous continuum formulation for the nonlinear elastic behavior of 2D materials[23]. By least-square fitting the DFT data before plastic region, fifteen elastic constants for the nonlinear continuum description of stanene are determined, as tabulated in Table 1 and Table S2. The closeness in $C_{11}$ and $C_{22}$ matches the near coincidence of the stress-strain responses of the two orthogonal loading modes in the linear-elastic region at small strains. Taking the numerical results obtained by the HSE06 functional as an example, the 2D Young's modulus of stanene is obtained by $E \approx E_\parallel = \frac{C_{11}^2 - C_{12}^2}{C_{11}} = 21.842$ N/m, based on plane stress condition. If the true stress - true strain curve is rather used to determine E, it is sensitive to the choice of onset point for nonlinearity, which could be ambiguous as illustrated in Figure S2.

The comparison of different exchange-correlational functionals in Table 2 reveals that UTS values of stanene are consistently within a narrow regime, confirming the accuracy of our calculations. Taking the averaged computational results based on hybrid functionals as the most accurate here, the UTS values are slightly over-predicted by the LDA functionals and are nearly the same as estimated by the two PBEsol functionals. From a rigorous perspective, the discussions in rest of this paper will be based on comparing the average mechanical properties values computed from using different functionals.

### 3.2 Comparison of Mechanical Properties and Ideal Strength

The DFT-calculated mechanical properties of stanene are compared to graphene and silicene in Table 3. The average 2D Young's modulus of stanene is computed to be 26.684 N/m, which is about half of silicene[43] and less than one-tenth of graphene[23,24]. A similar decrease in UTS also exists for stanene as opposed to graphene and silicene, demonstrating a trend of reducing stiffness and strength down the periodic table for Group-IV elements. This variation in mechanical properties could be attributed to the growing tendency of sp$^2$-sp$^3$ hybridization with increasing atomic radius and bond length going down Group-IV atoms. Higher bond length leads to less π-bond overlap and more sp$^3$ component, causing greater buckling ($\delta_{Sn} > \delta_{Ge} > \delta_{Si} > \delta_C = 0$). As a flat structure has much stronger covalent bonds formed by sp$^2$ hybridization, stanene, which has the least extent of sp$^2$, would have its π-bond mostly weakened as opposed to graphene and have the lowest in-plane stiffness among existing Group-IV monolayers. A visualization of the sp$^2$-sp$^3$ hybridization in relation to structural integrity can be found in Sec. 3.4.

Our DFT results suggest the ratio of Young's modulus to UTS is 6.519 for stanene measured along the zigzag direction and 7.341 along the armchair direction (Table 3), which are higher than the Griffith theoretical limit (UTS ≈ E/9). This underestimation by Griffith's criterion does not happen for silicene and graphene. When comparing to a more conservative estimate (~E/10) proposed by Cottrell[10] and based on the Polanyi-Orowan equation[8], the two lighter Group-IV monolayer materials do not show stress levels surpassing the threshold either. But it can already be seen the E/UTS ratio has become smaller going down the Group-IV list and, eventually, the breakdown of conventional estimates was found to happen at stanene. In the next section, phonon stability would be tested before making a definitive judgment on whether Griffith theoretical strength limit for uniaxial tension is indeed challenged in our case.

### 3.3 Phonon Instability

As discussed above, it seems from the stress-strain curve alone that stanene is stronger along the zigzag direction than armchair direction, and in both cases the UTS seemingly surpasses the Griffith and Cottrell strength limit. However, one needs to verify whether the stanene remains structurally stable upon reaching the maximum stress, as phonon instability[4,42] may intrude before the peak stress on the strain path.

The phonon dispersions plotted along high symmetrical points M', Γ, K', and M' for uniaxial armchair tension is shown in Figure 3. There are six phonon branches in total, three of which have an acoustic nature – the flexural acoustic (ZA), transverse acoustic (TA), and longitudinal acoustic (LA) branches. Separate phonon calculation of stanene at an undeformed state suggests that the acoustic branches are separated by a gap of ~48 cm$^{-1}$ below three optical branches (see Figure S3). It is evident from Figure 3a that stanene experienced phonon softening during deformation, and eventually, at a critical true strain ($\varepsilon_{xx}$) of 0.205, incurred negative (imaginary) frequencies near the Γ point. Analysis of the dynamical matrix shows that the soft mode at 0.205 armchair strain is related to the ZA mode. This strain corresponds to a position on the stress-strain curve right after the true strain at UTS (0.199), computed by adopting the LDA, suggesting that stanene will maintain vibrational stability at a theoretical strength level likely to surpass the Griffith limit when loaded in the armchair direction.

Under uniaxial tension along the zigzag direction, the onset of the imaginary frequencies is found to be at $\varepsilon_{yy} \approx 0.210$, with phonon soft mode of the ZA type, as shown in Figure 3b. This is prior to the strain (> 0.3) corresponding to the peak stress on a zigzag curve. Hence, the monolayer experiences phonon instability and the highest mechanical stress along the zigzag direction listed in Table 3 would not be readily achieved for stanene. This left the critical strain and stress to be about the same as that of armchair loading. Therefore, by taking overall consideration of both mechanical and phonon behavior, the final suggested achievable E-σ relation is determined to be $\sigma_{ideal} = E/7.4$ for perfect 2D stanene under both uniaxial tension directions. This implies, from a computational perspective, a possible surpass of the conventional ideal strength limit, where $\sigma_{ideal}$ falls between E/10~E/9 for solids at the continuum level.

### 3.4 Charge Density Analysis

The distinction between material responses to different uniaxial tensile strain can be manifested by employing simulated STM with a negative bias voltage, which probes occupied electron regions[44,45]. As discussed in Section 3.3, in contrast to perfectly flat graphene, the buckled shape of stanene is a direct result of $sp^2$-$sp^3$ hybridization. Figure 4a shows that Sn atoms and the associated electron clouds participating in forming the hybrid bond are in an alternating top and bottom position of the stanene basal plane, resulting in an arrangement of the Sn atoms belonging to the $D_{3d}^3$ point group. The overlapping between electron clouds in a buckled stanene is less significant than in graphene possessing a flat geometry attributed to pure $sp^2$ hybridization, which explains the drop in stiffness and strength as discussed in Sec. 3.2. The color of the bonding region qualitatively describes the extent of cohesion, and it is clear that upon loading the electrons accumulated in this region become less, indicating a weakening of the bond and a separation of adjacent atoms. Meanwhile, the Sn-Sn bond also experiences slight rotation and translation, which causes a decrease in the buckling distance of the monolayer stanene, as shown in Figure 4b.

It is found that atomic bonds which are parallel or make a small angle to the pulling direction are easier to break (Figure 4c). For stanene under uniaxial tension, $sp^2$-$sp^3$ bonds that are more aligned to the direction of pulling would break first while other bonds retain their less-aligned positions. In the case of equi-biaxial tension, the material quickly flattens out (Figure 4b) and all the bonds have an equal possibility to break, as evidenced by the same electron density around the hexagonal lattice (see Movie S1 for animations). It is worthwhile to note that the fracture pathways revealed by the simulated STM images and charge density analysis match the onset of the imaginary acoustic mode discussed Section 3.3, which further consolidates our findings in the mechanical behavior of stanene.

## 4. CONCLUSION

In summary, we studied the mechanical response of stanene under uniaxial and biaxial loading conditions using first-principles DFT calculations. Specifically, it was found that stanene could sustain up to ~20% deformation in its armchair and zigzag directions without losing either elastic or phonon stability. By fitting nonlinear continuum theory to obtained stress-strain curves, we evaluated the complete set of nonlinear elastic constants of stanene up to fifth order. A Young's modulus of 26.684 N/m was obtained for stanene. Our simulations were conducted with six different exchange-correlation functionals, and the calculated mechanical properties were found to be within a narrow range. This reflects the high chemical accuracy of our results and the relative insensitivity of mechanical response to the choice of DFT functional. Additionally, we explained through virtual STM and charge density analysis that the reducing stiffness and strength down the Group-IV 2D materials is owing to the increasing $sp^2$-$sp^3$ hybridization within the material structure.

Moreover, our theoretical calculations revealed the ideal strength of stanene at 0 K is about E/7.4, higher than the Griffith's and Cottrell's estimation of strength. However, the accuracy of strength predicted here could also depend upon any discrepancy in the atomic structure between theory and experiments. Therefore, it still needs to be confirmed by *in-situ* experiments on stanene for its mechanical property characterization. As is the case for graphene and $MoS_2$ where DFT-computed strengths have already been achieved experimentally, it is hoped that the breakdown of conventional strength limit in 2D materials as suggested here could possibly be realized in future.

## ACKNOWLEDGEMENTS

Financial support for this work was provided by the Natural Sciences and Engineering Research Council (NSERC) of Canada. The computational resources were provided by the WestGrid consortium through the Western Canada Research Grid and the SciNet consortium through the Compute Canada resource allocations.

**Note**: The authors declare no competing financial interest.

## REFERENCES

1  C. Lee, X. Wei, J. W. Kysar and J. Hone, *Science*, 2008, **321**, 385–388.
2  J. Wang, F. Sansoz, J. Huang, Y. Liu, S. Sun, Z. Zhang and S. X. Mao, *Nat. Commun.*, 2013, **4**, 1742.
3  J. Pokluda, M. Černý, P. Šandera and M. Šob, *J. Comput.-Aided Mater. Des.*, 2005, **11**, 1–28.
4  F. Liu, P. Ming and J. Li, *Phys. Rev. B*, 2007, **76**.
5  T. Zhu and J. Li, *Prog. Mater. Sci.*, 2010, **55**, 710–757.
6  A. A. Griffith, *Philos. Trans. R. Soc. Lond. Ser. A*, 1921, **221**, 163–198.
7  M. Polanyi, *Z. Für Phys.*, **7**, 323–327.
8  E. Orowan, *Rep. Prog. Phys.*, 1949, **12**, 185.
9  A. Kelly, *Strong solids*, Oxford, Clarendon Press, 1973., 1973.

10 A. H. Cottrell, *The Mechanical Properties of Matter*, Krieger Pub Co, Huntington, N.Y, 1981.
11 C. Deng and F. Sansoz, *ACS Nano*, 2009, **3**, 3001–3008.
12 S. M. Mirvakili, A. Pazukha, W. Sikkema, C. W. Sinclair, G. M. Spinks, R. H. Baughman and J. D. W. Madden, *Adv. Funct. Mater.*, 2013, **23**, 4311–4316.
13 Z. Shi and C. V. Singh, *Scr. Mater.*, 2016, **113**, 214–217.
14 J.-Y. Kim, D. Jang and J. R. Greer, *Acta Mater.*, 2010, **58**, 2355–2363.
15 J.-Y. Kim, D. Jang and J. R. Greer, *Int. J. Plast.*, 2012, **28**, 46–52.
16 R. Cao, Y. Deng and C. Deng, *Acta Mater.*, 2015, **86**, 15–22.
17 M. Liu, V. I. Artyukhov, H. Lee, F. Xu and B. I. Yakobson, *ACS Nano*, 2013, **7**, 10075–10082.
18 S. Bertolazzi, J. Brivio and A. Kis, *ACS Nano*, 2011, **5**, 9703–9709.
19 Y. Cai, G. Zhang and Y.-W. Zhang, *J. Am. Chem. Soc.*, 2014, **136**, 6269–6275.
20 H. Wang, Q. Li, Y. Gao, F. Miao, X.-F. Zhou and X. G. Wan, *New J. Phys.*, 2016, **18**, 073016.
21 Q. Peng, W. Ji and S. De, *Comput. Mater. Sci.*, 2012, **56**, 11–17.
22 Q. Peng, C. Liang, W. Ji and S. De, *Mech. Mater.*, 2013, **64**, 135–141.
23 X. Wei, B. Fragneaud, C. A. Marianetti and J. W. Kysar, *Phys. Rev. B*, 2009, **80**.
24 Q. Peng, C. Liang, W. Ji and S. De, *Phys Chem Chem Phys*, 2013, **15**, 2003–2011.
25 Q. Peng and S. De, *Nanoscale*, 2014, **6**, 12071–12079.
26 D. Roundy and M. L. Cohen, *Phys. Rev. B*, 2001, **64**, 212103.
27 J. Wang, Y. Xu and S.-C. Zhang, *Phys. Rev. B*, 2014, **90**, 054503.
28 H. Zhou, Y. Cai, G. Zhang and Y.-W. Zhang, *Phys. Rev. B*, 2016, **94**, 045423.
29 S.-C. Wu, G. Shan and B. Yan, *Phys. Rev. Lett.*, 2014, **113**, 256401.
30 L. Takahashi and K. Takahashi, *Phys. Chem. Chem. Phys.*, 2015, **17**, 21394–21396.
31 F. Zhu, W. Chen, Y. Xu, C. Gao, D. Guan, C. Liu, D. Qian, S.-C. Zhang and J. Jia, *Nat. Mater.*, 2015, **14**, 1020–1025.
32 J. Deng, A. Zhao and B. Wang, *33rd Int. Conf. Phys. Semicond.*, 2016.
33 S. Saxena, R. P. Chaudhary and S. Shukla, *Sci. Rep.*, 2016, **6**, 31073.
34 P. Giannozzi, S. Baroni, N. Bonini, M. Calandra, R. Car, C. Cavazzoni, Davide Ceresoli, G. L. Chiarotti, M. Cococcioni, I. Dabo, A. D. Corso, S. de Gironcoli, S. Fabris, G. Fratesi, R. Gebauer, U. Gerstmann, C. Gougoussis, Anton Kokalj, M. Lazzeri, L. Martin-Samos, N. Marzari, F. Mauri, R. Mazzarello, Stefano Paolini, A. Pasquarello, L. Paulatto, C. Sbraccia, S. Scandolo, G. Sclauzero, A. P. Seitsonen, A. Smogunov, P. Umari and R. M. Wentzcovitch, *J. Phys. Condens. Matter*, 2009, **21**, 395502.
35 J. P. Perdew, A. Ruzsinszky, G. I. Csonka, O. A. Vydrov, G. E. Scuseria, L. A. Constantin, X. Zhou and K. Burke, *Phys. Rev. Lett.*, 2008, **100**, 136406.
36 N. Troullier and J. L. Martins, *Phys. Rev. B*, 1991, **43**, 1993–2006.
37 C. Hartwigsen, S. Goedecker and J. Hutter, *Phys. Rev. B*, 1998, **58**, 3641–3662.
38 B. van den Broek, M. Houssa, E. Scalise, G. Pourtois, V. V. Afanas'ev and A. Stesmans, *2D Mater.*, 2014, **1**, 021004.
39 Y. Xu, B. Yan, H.-J. Zhang, J. Wang, G. Xu, P. Tang, W. Duan and S.-C. Zhang, *Phys. Rev. Lett.*, 2013, **111**, 136804.
40 D. J. Bonet and D. R. D. Wood, *Nonlinear Continuum Mechanics for Finite Element Analysis*, Cambridge University Press, Cambridge, UK ; New York, 2 edition., 2008.
41 J. F. Nye, *Physical properties of crystals : their representation by tensors and matrices*, Oxford [Oxfordshire] : Clarendon Press ; New York : Oxford University Press, 1985., 1985.

42 S. Baroni, S. de Gironcoli, A. Dal Corso and P. Giannozzi, *Rev. Mod. Phys.*, 2001, **73**, 515–562.
43 Q. Peng, X. Wen and S. De, *RSC Adv.*, 2013, **3**, 13772.
44 D. R. Vij, *Handbook of applied solid state spectroscopy*, New York : Springer, c2006., 2006.
45 Z. Hou, X. Wang, T. Ikeda, K. Terakura, M. Oshima and M. Kakimoto, *Phys. Rev. B*, 2013, **87**, 165401.

# Tables

Table 1. Nonzero second- and higher-order elastic constants (in Voigt notation and unit N/m) tabulated below for the HSE06 functional.

| $2^{nd}$-order | $3^{rd}$-order | $4^{th}$-order | $5^{th}$-order |
|---|---|---|---|
| $C_{11}$ = 26.64 | $C_{111}$ = -186.6 | $C_{1111}$ = 597.2 | $C_{11111}$ = -173.3 |
| $C_{12}$ = 11.306 | $C_{112}$ = -78.4 | $C_{1112}$ = 484.5 | $C_{11112}$ = -2206.2 |
| $C_{22}$ = 26.47 | $C_{222}$ = -226.9 | $C_{1122}$ = 172.4 | $C_{11122}$ = -3804.1 |
| | | $C_{2222}$ = 1590 | $C_{12222}$ = -7560.6 |
| | | | $C_{22222}$ = -7188.7 |

Table 2. Comparison of UTS and 2D Young's modulus (E) of the three loading modes obtained by adopting six different functionals. The complete data set for elastic constants can be found in Table S1 of Supplementary Materials.

| | *Unit: N/m* | PBEsol (RRKJ) | PBEsol (PAW) | PW-LDA | HSE06 | PBE0 | B3LYP |
|---|---|---|---|---|---|---|---|
| | Armchair | 3.656 | 3.718 | 4.076 | 2.919 | 3.821 | 3.622 |
| UTS | Zigzag | 4.084 | 4.097 | 4.520 | 3.380 | 4.224 | 3.930 |
| | Biaxial | 3.318 | 3.361 | 3.670 | 3.318 | 3.333 | 3.186 |
| E | Armchair, $E_\parallel$ | 24.448 | 26.600 | 29.865 | 21.842 | 26.470 | 30.880 |
| | Zigzag, $E_\perp$ | 23.897 | 26.685 | 29.637 | 21.637 | 27.670 | 30.620 |

Table 3. Comparison of the calculated mechanical properties of stanene, silicene, and graphene.

| | *Stress unit: N/m* | Stanene (averaged by 6 functionals) | Silicene[43] | Graphene[23] | Graphene[24] |
|---|---|---|---|---|---|
| | Young's modulus, E (DFT) | 26.684 | 63.8 | 348 | 340.8 |
| | Theoretical strength limit, E/10-E/9 (Griffith and others[6,10]) | 2.668-2.965 | 6.38-7.09 | 34.8-38.67 | 34.08-37.87 |
| UTS (DFT) | Armchair | 3.635 | 6.0 | 29.5 | 28.6 |
| | Zigzag | 4.093 | 5.9 | 31.4 | 30.4 |
| E/UTS ratio | Armchair | 7.341 | 10.3 | 11.8 | 11.9 |
| | Zigzag | 6.519 | 10.8 | 11.1 | 11.2 |
| | Poisson ratio $\nu$ | 0.434 | 0.325 | 0.169 | 0.178 |

Figure 1. The structure of stanene. (a) Ball-stick display of stanene superimposed on a virtual STM image showing the top view of the honeycomb lattice structure. A six-atom supercell for stress-strain calculations is encircled. (b) Perspective view of undeformed stanene showing the buckled distance, bond length, and bond angle at the undeformed state. (c) The two-atom primitive cell for phonon calculations and (d) the associated reciprocal space with the first Brillouin zone.

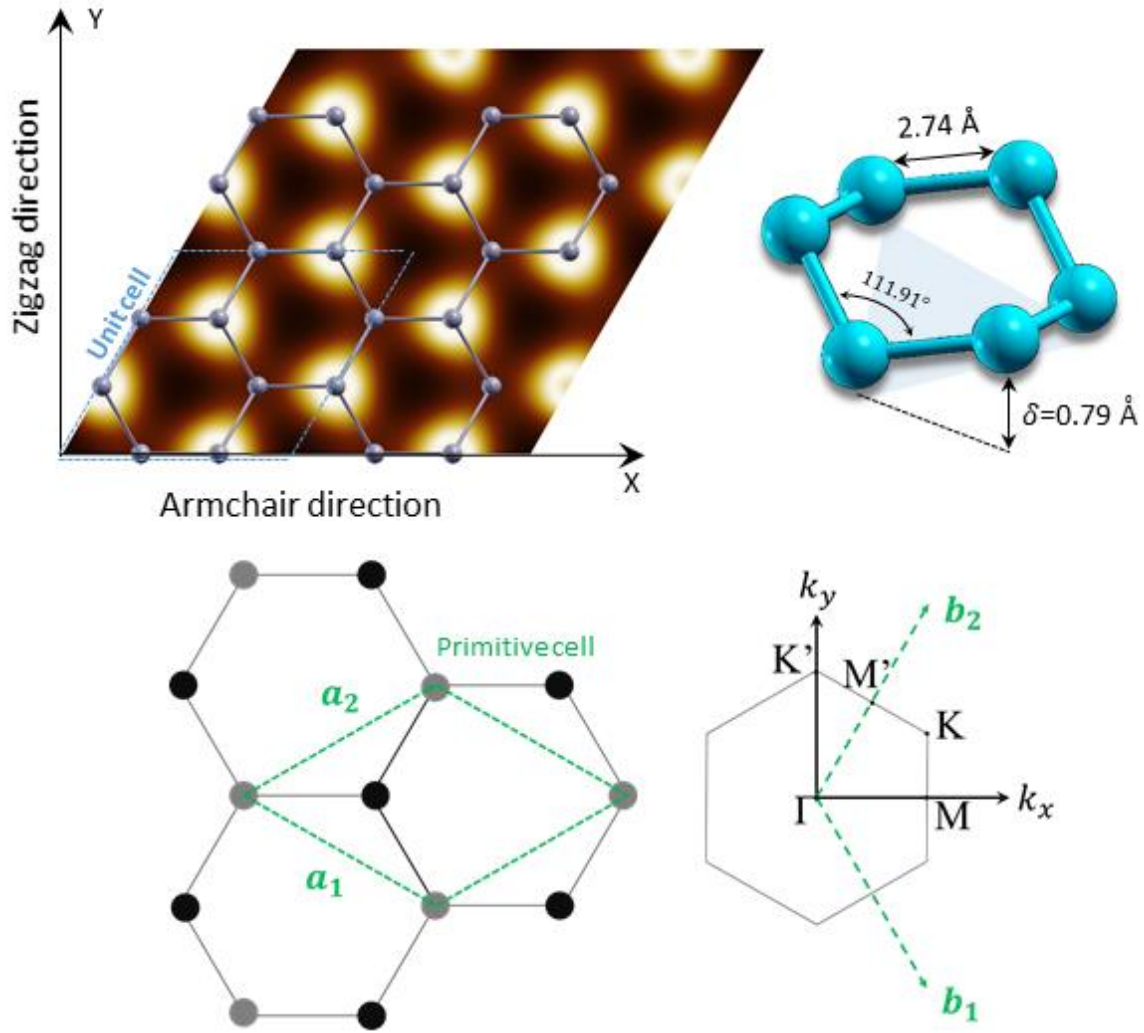

Figure 2. (a) The stress-strain response of stanene. Quantities are plotted in PK2 stress and Lagrangian strain. The lines indicate least-squares fit to the DFT data. (b) The same data converted to true stress and true strain. For demonstration purpose, only the results obtained by PBEsol (RRKJ) calculations are shown here.

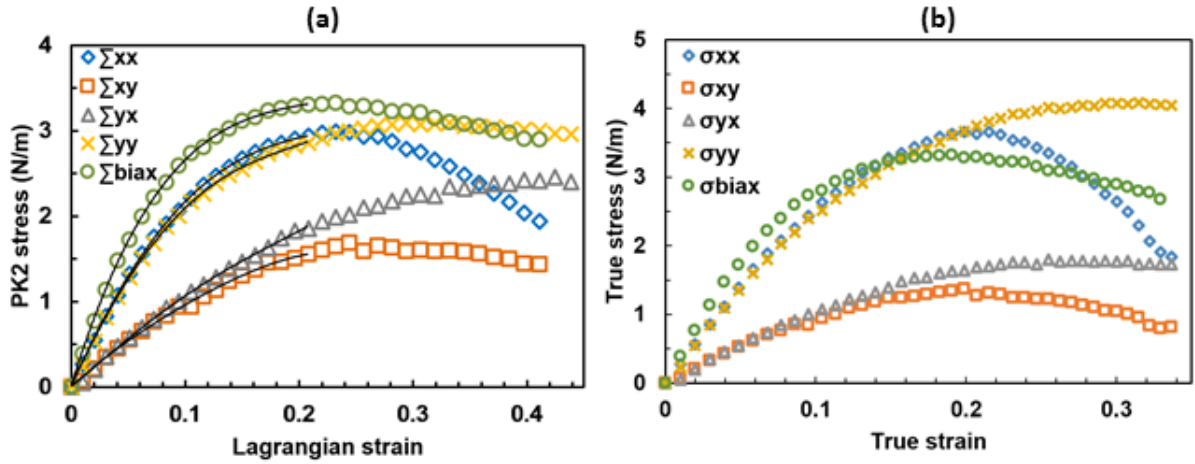

Figure 3. Phonon dispersions for stanene under uniaxial tension. The onsets of imaginary frequencies under (a) armchair and (b) zigzag tension are indicated.

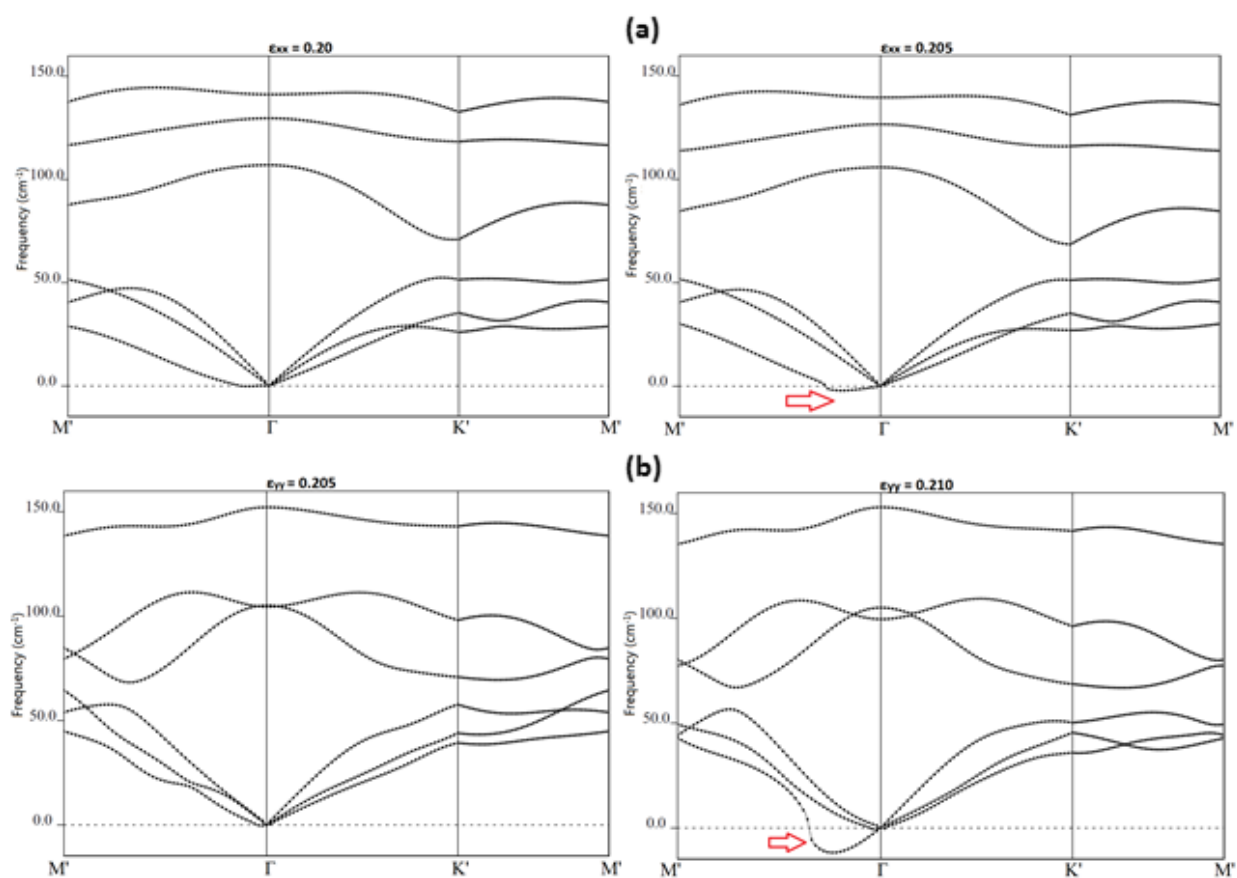

Figure 4. (a) Simulated STM images and charge-density plots illustrating charge distribution between adjacent Sn atoms at undeformed state and after bond breakage. (b) Decreasing buckled distance for stanene under tension. (c) Simulated STM images showing the alternation of electron distribution during tensile tests. Out-of-plane charge dispersion is also included for each case. White × indicates the most probable position for bond breakage in a supercell if the strain is sufficiently high.

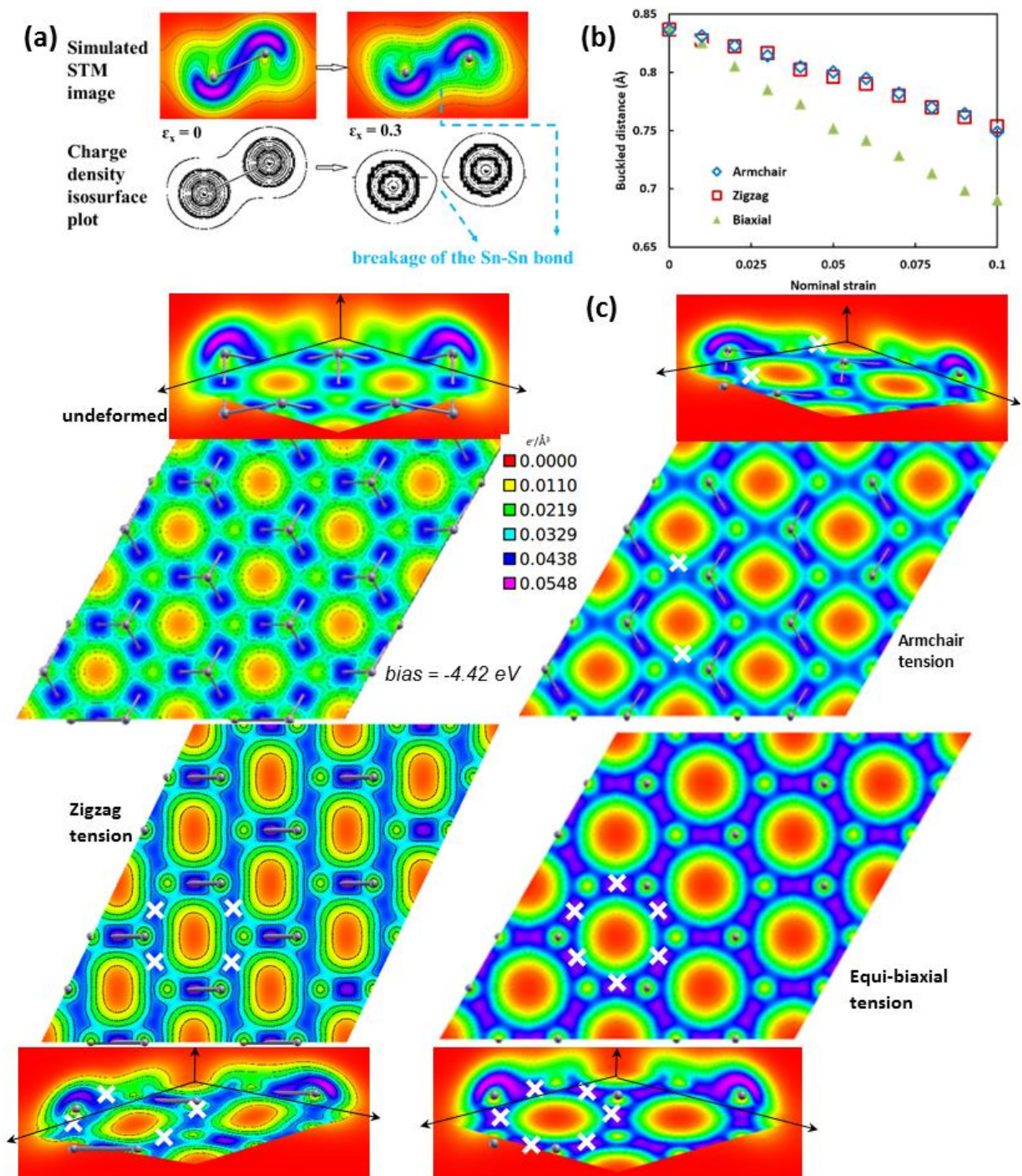